\documentclass[12pt,a4paper]{article}
\input{epsf}
\title{Fluid dynamical description of the chiral transition \footnote{To be published in the Proceedings of the International Workshop on Applicability of Relativistic Hydrodynamical Models in Heavy Ion Physics held in Trento, May 12-16, 1997}}

\author{{I.N. Mishustin$^{1,2}$, J.A. Pedersen$^{1}$ and O. Scavenius$^{1}$} \\
{\it $^{1}$The Niels Bohr Institute, University of Copenhagen,}\\
{\it Blegdamsvej 17, DK-2100 Copenhagen \O , Denmark} \\ 
{\it $^{2}$The Kurchatov Institute, Russian Scientific Center,}\\
{\it Moscow, 123182 Russia}}

\date{June 25, 1997}

\begin{document}

\maketitle

\begin{abstract}
We investigate the dynamics of the chiral transition in an expanding quark-anti-quark plasma. 
The calculations are made within a linear $\sigma$-model with explicit quark and antiquark degrees of freedom. 
We solve numerically the classical equations of motion for chiral fields coupled to the fluid dynamical equations for the plasma.
Fast initial growth and strong oscillations of the chiral field and strong amplification of long wavelength modes of the pion field are observed in the course of the chiral transition. 
\end{abstract}

{\em Introduction.--} It is commonly believed that color deconfinement and chiral symmetry restoration take place at early stages of ultra-relativistic heavy-ion collisions.
 At intermediate stages of the reaction the quark-gluon plasma may be formed and evolve through the states close to thermodynamical equilibrium.
 However, at later stages of the expansion the transition to the hadronic phase with broken chiral symmetry should take place.
 The breakdown of chiral symmetry will possibly lead to such interesting phenomena as formation of disoriented chiral condensates (DCCs) and classical pion fields, as well as clustering of quarks and antiquarks.
 These phenomena were studied recently in many publications [1-15], using QCD motivated effective models, such as the linear and nonlinear $\sigma$-models and the Nambu\---Jona-Lasinio (NJL) model.
 Of course, these models have some significant shortcomings, e.g. they do not possess color confinement and the NJL-model is non-renormalizable. 
The key point is, however, that these models obey the same chiral symmetry as the QCD Lagrangian.

In most applications of the $\sigma$-model the quark degrees of freedom are disregarded (see e.g.[2-9,15]).
The inclusion of quarks \cite{mishu,mismocs,ms}  makes it possible to study the hadronization process. 
 In this letter, we construct a relativistic mean field fluid dynamical model based on the linear $\sigma$-model assuming that the quarks and antiquarks are in local thermodynamical equilibrium before, during, and after the chiral transition. 
This is an idealization which certainly cannot be true at later stages of the expansion.
On the other hand, in \cite{mishu,mismocs,ms} we have considered another limiting case assuming a collisionless approximation for quarks and antiquarks.
Below, we assume that the expansion can be well approximated by a spherical scaling solution at this stage.

{\em The linear $\sigma$-model.--} 
To describe the complex dynamics of the chiral transition we adapt the linear $\sigma$-model where quarks and antiquarks interact with the classical $\sigma$ and $\vec\pi$ fields.
Together, the scalar field $\sigma$ and the pion field $\vec{\pi} =(\pi _{1},\pi _{2},\pi _{3})$ form a chiral field $\Phi =(\sigma,\vec{\pi})$.
The equations of motion obtained from the linear $\sigma$-model Lagrangian are: 
\begin{displaymath}
\partial _{\mu}\partial ^{\mu}\sigma (x)+\lambda^{2}[\sigma ^{2} (x)+\vec{\pi} ^{2} (x)-v^{2}]\sigma(x)-H=-g\rho_{s}(x),\\
\end{displaymath}
\begin{equation}
\partial _{\mu}\partial ^{\mu}\vec{\pi} (x)+\lambda^{2}[\sigma ^{2} (x)+\vec{\pi}^{2} (x)-v^{2}]\vec{\pi}(x)=-g\vec{\rho}_{p}(x),
\label{sigmab}
\end{equation}     
where $\rho _{s}=\langle\overline{q}q\rangle$ and $\vec{\rho}_{p}=i\langle\overline{q}\gamma _{5}\vec{\tau} q\rangle$ are the scalar and pseudoscalar densities generated by valence $u$ and $d$ quarks.
 They are determined by the interaction of quarks with meson fields. 
As shown in \cite{mishu} these densities can be expressed as $\rho _{s}(x)=g\sigma (x)a(x)$ and $\vec{\rho}_{p}=g\vec{\pi} (x)a(x)$, with the scaling function 
\begin{equation}
a(x)=\frac{\nu _{q}}{(2\pi )^{3}}\int \frac{d^{3}p}{\sqrt{{\bf p}^{2}+m^{2}(x)}}[n_{q}(x,{\bf p})+n_{\overline{q}}(x,{\bf p})].
\label{af}
\end{equation}
Here $\nu_{q}=2\times2\times3=12$ is the degeneracy factor of quarks and $m(x)$ is the constituent quark mass which is determined self-consistently through the meson fields,
\begin{equation}
m^{2}(x)=g^{2}[\sigma ^{2}(x)+\vec{\pi}^{2}(x)].
\label{m}
\end{equation} 
In eq.(\ref{af}) $n_{q}$ and $n_{\overline{q}}$ are the quark and antiquark occupation numbers, which in thermal equilibrium are given by the Fermi-Dirac distributions. 
This linear $\sigma$-model is approximately invariant under chiral $SU_{L}(2) \otimes SU_{R}(2)$ transformations if the explicit symmetry breaking term $H\sigma $ is small.
 The parameters are chosen in such a way that the chiral symmetry is spontaneously broken in the vacuum so that the expectation values of the meson fields are $\langle\sigma\rangle ={\it f}_{\pi}$ and $\langle\vec{\pi}\rangle =0$, where ${\it f}_{\pi}=93$ MeV is the pion decay constant. 
The constant $H$ is fixed by the PCAC relation that gives $H=f_{\pi}m_{\pi}^{2}$, where $m_{\pi}=138$ MeV is the pion mass.
 Then one finds $v^{2}=f^{2}_{\pi}-\frac{m^{2}_{\pi}}{\lambda ^{2}}$. 
$\lambda ^{2}$ is determined by the sigma mass, ${m_{\sigma}}^{2}=2\lambda ^{2}f^{2}_{\pi}+m^{2}_{\pi}$, which we set to 600 MeV yielding $\lambda ^{2} \approx 20$. 
The coupling constant $g$ is fixed by the requirement that the constituent quark mass in vacuum, $m_{vac}=gf_{\pi}$, is about $1/3$ of the nucleon mass, giving
$g \approx 3.3$.  
With these parameters a chiral phase transition is predicted at $T_{c} \approx 132$ MeV \cite{mismocs}.

{\em Fluid dynamical equations.--}
The relativistic Vlasov equation consistent with the linear $\sigma$-model can be written as:
\begin{equation}
\left[p_{\mu}\frac{\partial}{\partial x_{\mu}}+\frac{1}{2}\frac{\partial m^{2}(x)}{\partial x_{\mu}} \cdot \frac{\partial}{\partial p_{\mu}}\right]{\it f}(x,p)=I_{coll},
\label{vlasov}
\end{equation} 
where $I_{coll}$ is the collision integral and $f(x,p)$ the scalar part of the fermion distribution function. 
Applying the moment expansion method and assuming local thermal equilibrium ($I_{coll}=0$) one arrives at the fluid dynamical equations of the form:
\begin{equation}
\partial_{\mu}T^{\mu\nu}+\frac{a}{2}\partial_{\mu}m^{2}=0,
\label{eqf}
\end{equation}
The energy-momentum tensor can be expressed in terms of the rest frame energy density $\mathcal{E}$ and pressure $P$ of quarks as:
\begin{equation}
T^{\mu\nu}=(\mathcal{E}+P)u^{\mu}u^{\nu}-Pg^{\mu\nu}.
\label{emt}
\end{equation}
where $u^{\mu}$ is the collective 4-velocity.
$\mathcal{E}$ and $P$ read:
\begin{equation}
\mathcal{E}=\nu_{q}\int\frac{d^{3}p}{(2\pi)^{3}}\sqrt{p^{2}+m^{2}}[n_{q}(p)+n_{\overline{q}}(p)],
\label{ed}
\end{equation}
and
\begin{equation}
P=\frac{\nu_{q}}{3}\int\frac{d^{3}p}{(2\pi)^{3}}\frac{p^{2}}{\sqrt{p^{2}+m^{2}}}[n_{q}(p)+n_{\overline{q}}(p)],
\label{p}
\end{equation}
Here $m$ is given by eq.(\ref{m}), and $n_{q}(p)$ and $n_{\overline{q}}(p)$ are the quark and antiquark occupation numbers:
\begin{equation}
n_{q}(p)=\left[\exp\left(\frac{\sqrt{p^{2}+m^{2}}-\mu}{T}\right)+1 \right]^{-1},\mbox{ } n_{\overline{q}}(p)=n_{q}(p, \mu\rightarrow -\mu). 
\label{occu}
\end{equation} 
In this way (\ref{sigmab}), (\ref{m}) and (\ref{eqf}) constitute a self-consistent set of differential equations which can be solved numerically for meson fields and quark distributions. Below we put the quark chemical potential to zero, $\mu$=0 (baryon-free plasma).

\begin{figure}
\begin{center}
\mbox{
\epsfxsize=16cm
\epsffile{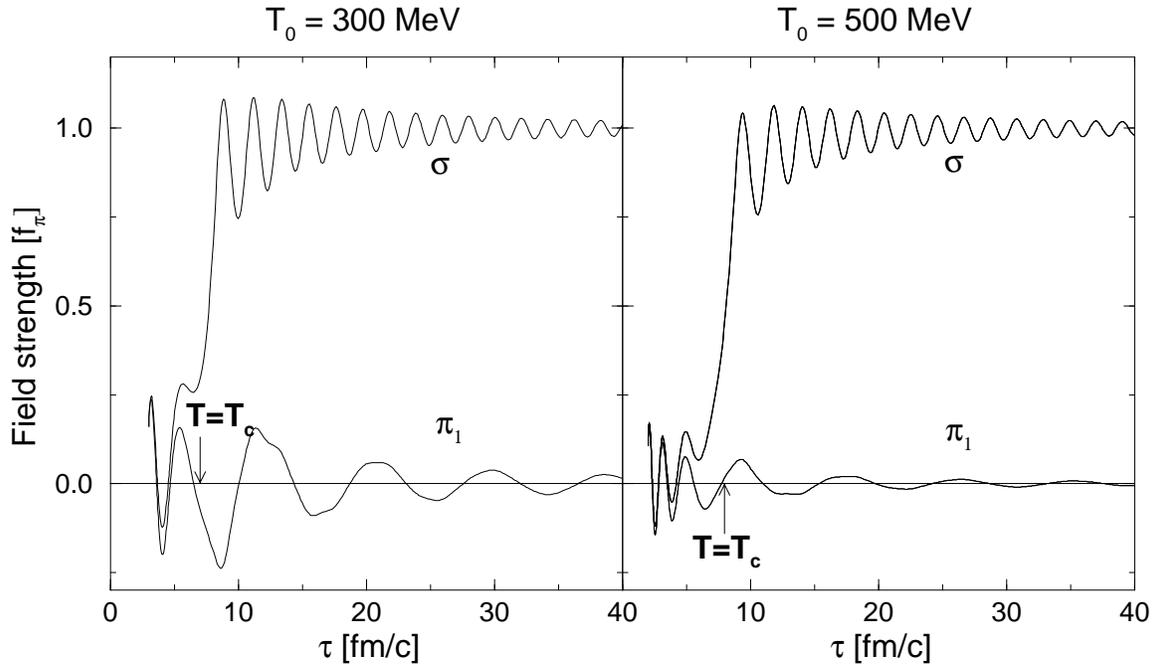}
}
\caption{Meson fields in units of $f_{\pi}$ as functions of proper time shown for two different initial temperatures. 
The initial conditions for the left figure are $\tau_{0}$ = 3 fm/c, T($\tau_{0}$) = 300 MeV, $\Phi(\tau_{0})$=(15,15,0,0) MeV and $(\partial\Phi/\partial\tau)_{\tau=\tau_{0}}$=(15000,15000,0,0) MeV/fm. 
For the right figure, $\tau_{0}$ = 2 fm/c, T($\tau_{0}$) = 500 MeV, $\Phi(\tau_{0})$=(10,10,0,0) MeV and $(\partial\Phi/\partial\tau)_{\tau=\tau_{0}}$=(20000,20000,0,0) MeV/fm.}
\label{fieldfig}
\end{center}
\end{figure}

In this letter we assume spherical scaling expansion: $u^{\mu}=x^{\mu}/\tau$ and $\tau=\sqrt{t^{2}-\vec{r}^{2}}$. 
All quantities now depend on $\tau$ only: $\sigma (\tau)$, $\vec{\pi}(\tau)$, $m(\tau)$, $\mathcal{E}(m(\tau),T(\tau))$ and $P(m(\tau),T(\tau))$. 
The d'Alembertian appearing in eqs.(\ref{sigmab}) is reduced to $\partial_{\mu}\partial^{\mu}\longrightarrow\frac{d^{2}}{d\tau^{2}}+\frac{3}{\tau}\frac{d}{d\tau}$. 
Moreover, eq.(\ref{eqf}) reads:
\begin{equation}
\frac{d\mathcal{E}}{d\tau}+3\frac{\mathcal{E}+P}{\tau}+\frac{a}{2}\frac{dm^{2}}{d\tau}=0.
\label{seqf}
\end{equation}
By introducing the dimensionless temperature $\theta=\frac{T}{m}$ this equation can be represented as
\begin{equation}
\frac{d\mathcal{E}}{d\tau}=\frac{\partial\mathcal{E}}{\partial m^{2}}\frac{dm^{2}}{d\tau}+\frac{\partial\mathcal{E}}{\partial\theta}\frac{d\theta}{d\tau}.
\label{enexp}
\end{equation}
$\theta$ can now be found by solving an equation of the form
\begin{equation}
\frac{1}{\theta^{2}}\frac{d\theta}{d\tau}=F_{1}(\theta)\frac{d\ln m^{2}}{d\tau}+\frac{F_2(\theta)}{\tau},
\label{thetaeq}
\end{equation}
where $F_{1}(\theta)$ and $F_{2}(\theta)$ are known functions.
One can easily find analytical solutions in two limiting cases. In the massless (ultrarelativistic) limit ($\theta=\frac{T}{m} \gg 1$), equation (\ref{thetaeq}) reduces to
\begin{equation}
\frac{1}{\theta^{2}}\frac{d\theta}{d\tau}=-\frac{1}{\theta\tau} \Longrightarrow \frac{d\theta}{d\tau}=-\frac{\theta}{\tau},
\label{massless}
\end{equation}
which yields 
\begin{equation}
T=\frac{\tau_0}{\tau}T_0.
\label{TtauMassless}
\end{equation}
In the Boltzmann (nonrelativistic) limit ($\theta=\frac{T}{m} \ll 1$), on the other hand, one obtains a logarithmic behavior,
\begin{equation}
\frac{1}{\theta^{2}}\frac{d\theta}{d\tau}=-\frac{3}{\tau}(1-\frac{1}{2}\theta) \Longrightarrow \frac{1}{\theta_0}-\frac{1}{\theta}=-3\ln\frac{\tau}{\tau_0},
\label{Boltzmann}
\end{equation}
which can be rearranged to give an expression for T as a function of $\tau$,
\begin{equation}
T(\tau)=\frac{m}{\frac{m}{T_0}+3\ln\frac{\tau}{\tau_0}}.
\label{TtauBoltzmann}
\end{equation}

{\em Initial conditions.--} In the chirally symmetric phase, the mean values of meson fields are close to zero.
We take initial values of fields and their derivatives from the Gaussian distributions around the mean values.
The probability distribution of field fluctuations $\delta \sigma$ and $\delta \vec{\pi}$ can be expressed through the change of the thermodynamical potential $\Omega$:

\begin{equation}
W(\delta\sigma,\delta\vec{\pi}) \sim \exp\left[-\frac{\Delta \Omega V}{T}\right].
\label{prob}
\end{equation}
To second order in field fluctuations, the expansion of the thermodynamical potential yields \cite{mismocs}
\begin{equation}
\Delta\Omega(\delta\sigma,\delta\vec{\pi})=\left[g^{2}\nu_{q}\frac{T^{2}}{12}-\lambda^{2}f_{\pi}^{2}\right]\frac{(\delta\sigma)^{2}+(\delta\vec{\pi})^{2}}{2}=\frac{g^{2}\nu_{q}}{12}(T^{2}-T_{c}^{2})\frac{(\delta\sigma)^{2}+(\delta\vec{\pi})^{2}}{2}.
\label{omegaexp}
\end{equation}
Thus, the initial fluctuations of the chiral field can be chosen within a sphere in isospace with radius equal to the variance of the Gaussian distribution,
\begin{equation}
\sigma_{var}=\frac{12 T}{g^{2}\nu_{q}(T^{2}-T_{c}^{2})V}.
\label{var}
\end{equation}

{\em Numerical results.--} We ran the simulation for several sets of randomly oriented initial conditions taken at the initial time $\tau_{0}$ and temperature $T=T(\tau_{0})$.
The time $\tau_{c}$ at which the temperature drops below the critical temperature $T_{c}$ depends on the initial values of $\tau_{0}$ and $T_{0}$. 
We have observed the following common trends. 
At low values of $\tau<\tau_{c}$, i.e. at high temperatures corresponding to the symmetric phase, the $\sigma$ and the $\pi_{i}$ fields are essentially identical, and they therefore oscillate with the same frequency and more or less the same amplitude (depending on how the energy is distributed between $d\sigma/d\tau$ and $d\pi_{i}/d\tau$). 

However, when the temperature approaches its critical value around $\tau_{c}$ = 7 fm/c and chiral transition starts, the field dynamics change. 
The $\sigma$ field rises rapidly, within 2-3 fm/c, and comes close to the vacuum value. After the transition, it oscillates around its vacuum value, $f_{\pi}$.
The $\pi_{i}$ components continue to oscillate around zero, but with decreasing frequency. 
Whereas the $\sigma$ field oscillates with approximately the same period ($\sim$ 2 fm/c) as before the transition, the $\pi_{i}$ oscillations slow down and show longer periods ($\sim$ 9 fm/c).
This is easy to understand because in normal vacuum the pion mass is about 4 times smaller than the sigma mass.
With the given initial values, the pion fields die out rather quickly (particularly for high initial temperatures).

\begin{figure}
\begin{center}
\mbox{
\epsfxsize=16cm
\epsffile{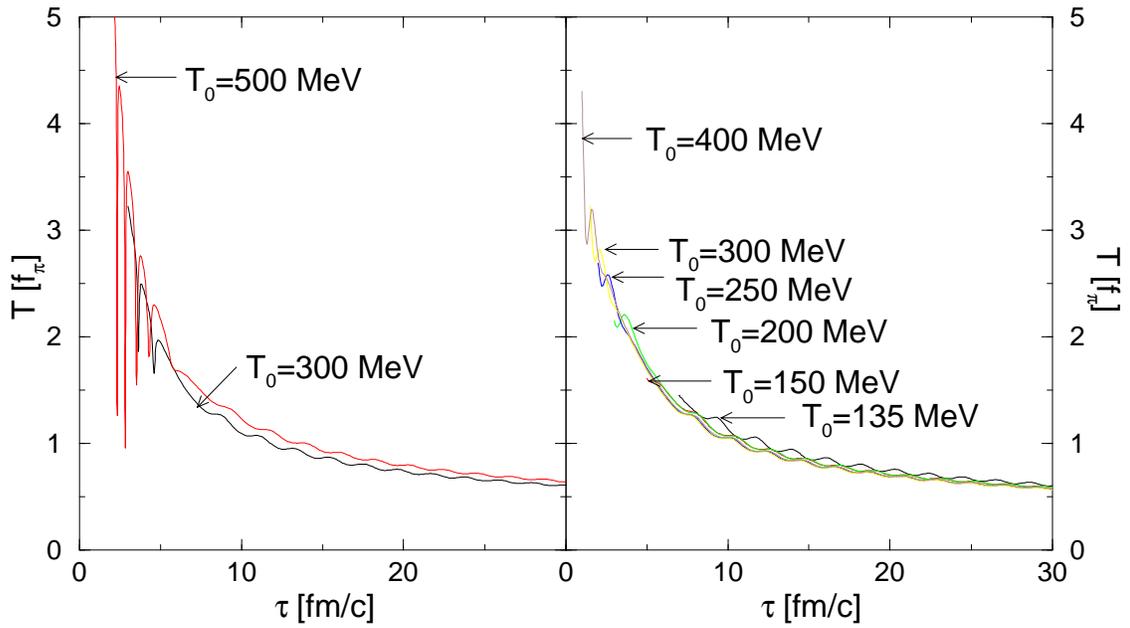}
}
\caption{Temperature as function of proper time for different initial temperatures. Initial conditions of left figure as in Figure \ref{fieldfig}. 
The sharp dips are purely numerical artifacts.
On the right, initial field fluctuations are distributed more evenly between the four chiral degrees of freedom.}
\label{tempfig}
\end{center}
\end{figure}

\begin{figure}
\begin{center}
\mbox{
\epsfxsize=16cm
\epsffile{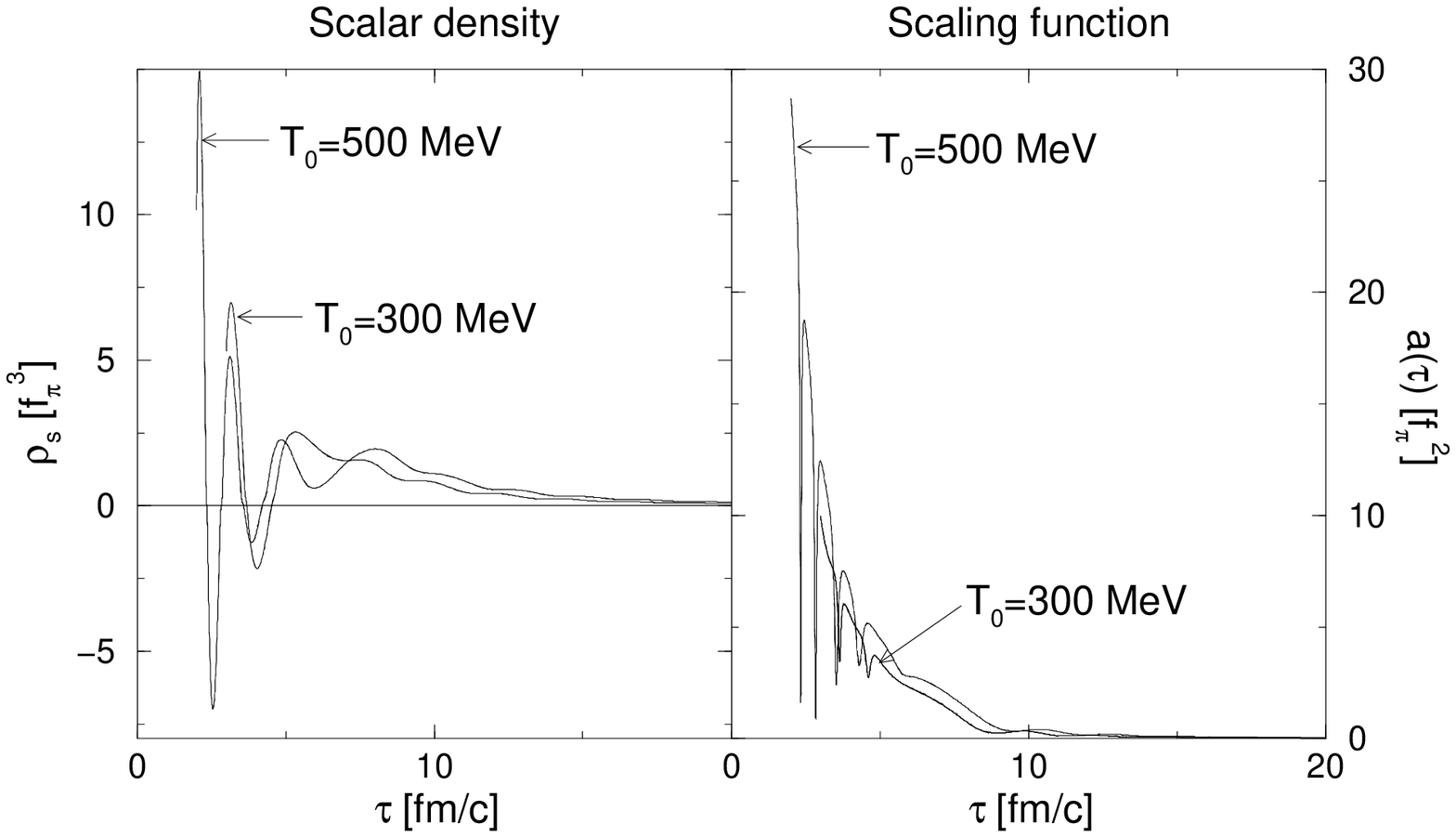}
}
\caption{Scalar density $\rho_{s}(\tau)$ and scaling function $a(\tau)$. 
Initial conditions as in figure \ref{fieldfig}. 
The sharp dips in the scaling function are purely numerical artifacts.}
\label{scalfig}
\end{center}
\end{figure}

In Figure \ref{tempfig} we show the temperature evolution for several initial temperatures, mainly to point out the uniformity of the evolution, which is to be expected when in each case we choose $\tau_{0}$ so that $T=T_{c}=132$ MeV is reached at around $\tau_{c}=7$ fm/c. 
Initially, the temperature oscillates around a decreasing curve which falls off like in the massless limit, as in eq.(\ref{TtauMassless}). 
But after the transition it slows down as the constituent mass rises, without quite reaching the behavior of the Boltzmann limit.
The left figure depicts the temperature curves for the two initial conditions shown in figure \ref{fieldfig}.

The right part of Figure \ref{tempfig} shows the evolution from several different initial states. 
The curves in the right frame don't exhibit the numerical dips of the $T_{0}$=500 MeV curve on the left because we have distributed the initial field strength and energy more evenly among all four chiral degrees of freedom.  
The curves overlap because we have chosen $\tau_{0}$ so that in each case the temperature reaches its critical value around $\tau_{c}$ = 7 fm/c. 
This figure emphasizes the universality of the evolution.

\begin{figure}
\begin{center}
\mbox{
\epsfxsize=14cm
\epsffile{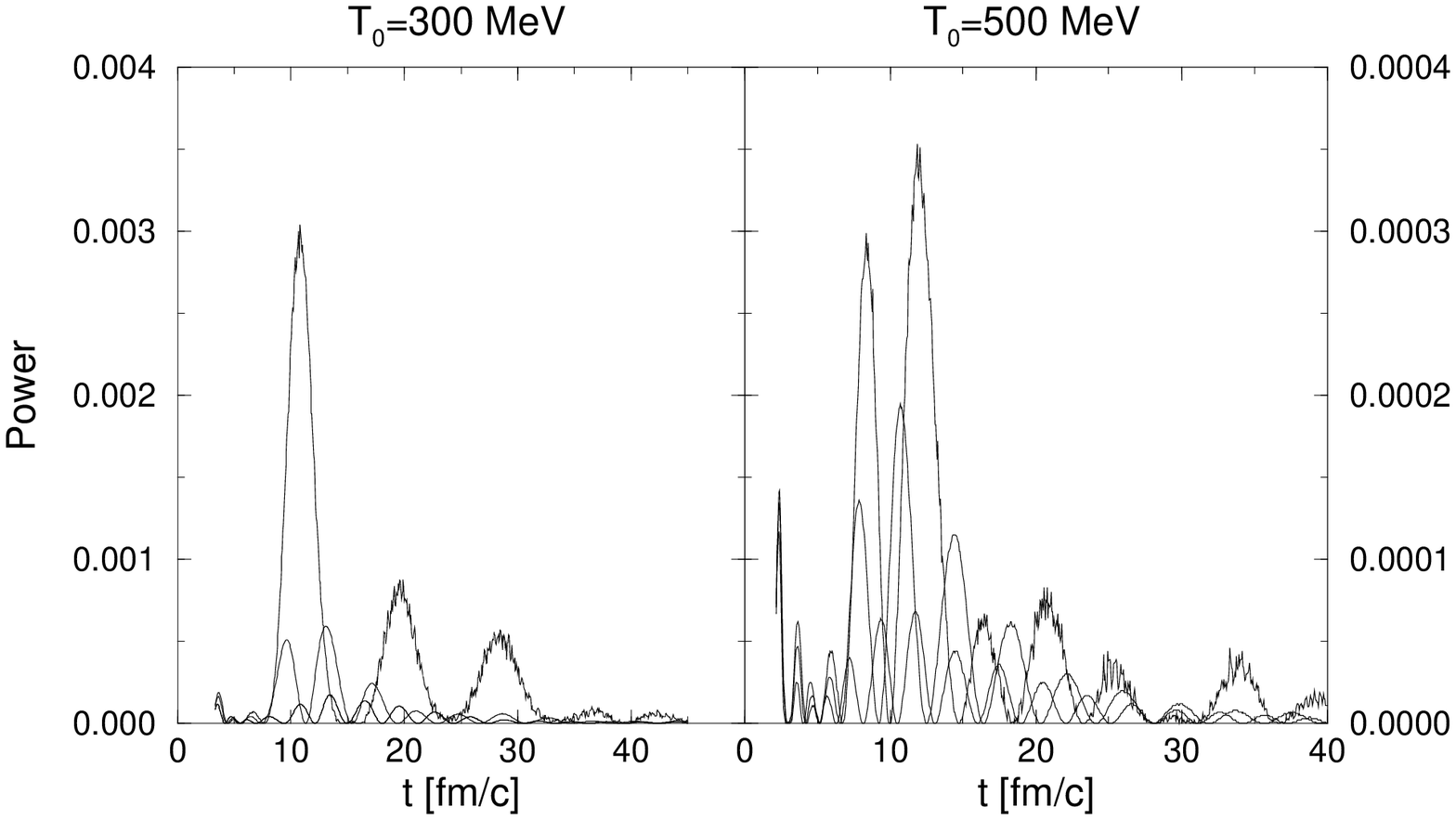}
}
\caption{The power of the modes: $k=10,80,150$ MeV translated into the $(t,r)$ space. 
They follow the rule of the smaller the mode the larger the amplitude }
\label{powerfig}
\end{center}
\end{figure}

Figure \ref{scalfig} shows the scalar density $\rho_{s}=ga(\tau)\sigma(\tau)$ and the scaling function $a(\tau)$ (as defined in eq.(\ref{af}), but now as a function of $\tau$), for the two scenarios considered in Figure \ref{fieldfig}. 
For an initial temperature $T_{0}=500$ MeV, $a(\tau)$ starts out at about twice as large a value as for $T_{0}=300$ MeV, but once it reaches the starting point of the $T=300$ MeV curve, it has decreased and they evolve close to each other.
From Figure \ref{scalfig} one can conclude that the scalar density of quarks and antiquarks drops out very rapidly after the transition temperarure. 
At later times the meson field dynamics is very little disturbed by the presence of quarks and antiquarks.

We have calculated the power spectrum of the first component of the pion field for the modes with wave numbers $k=10,80,150$.
In the present context the power spectrum was first discussed in \cite{rajg}.
It is defined as
\begin{equation}
S(t,k)=\left|\frac{4\pi}{k}\int^{R}_{0}rdr\sin(kr)\pi(t,r)\right|^{2},
\end{equation}
where $R=\sqrt{t^{2}-\tau^{2}}$ is the spatial radius of the system at time $t$.
The initial conditions are the same as in Figure \ref{fieldfig}. 
In the case of an initial temperature of $300$ MeV, one clearly observes the strong amplification of the low momentum mode as found in many other works \cite{rajg,sg,many3}. 
But we see that in the present case, partly due to the 3D expansion, the amplification of the pion field is weaker than found previously. 
Also, the response of the system depends sensitively on the initial conditions.This is clearly demonstrated by the power spectrum for $T=500$ MeV, where the response of the low momentum mode is reduced by an order of magnitude compared to the case of $300$ MeV. 
Clearly the field fluctuations will grow as the system approaches the critical temperature, and then they will be further amplified in the course of the chiral transition.
In the present schematic model this effect is not taken into account because we define the initial fluctuations deep inside the symmetric phase and then propagate them through the chiral transition.

{\em Summary.--} 
We have investigated the dynamics of the chiral transition in expanding quark-antiquark plasma.
The dynamics of chiral symmetry breaking and the time evolution of the chiral field were studied within a linear $\sigma$-model in a relativistic mean field fluid dynamical approach. 
This approach made it possible to study the evolution of chiral fields even before the chiral transition. 
Fast initial growth and strong oscillations of the chiral field and strong amplification of long wavelength modes of the pion field were observed. 
We are working on 3D simulations and taking dissipation into account.   
 \\
 \\
{\bf {\it Acknowledgement.--}} The authors thank R.F. Bedaque, J. Bjorken, D. Boyanovsky, T.S. Bir{\' o}, J.P. Bondorf, J. Borg, L.P. Csernai, A.D. Jackson, S. Klevansky, {\' A}. M{\' o}csy, J. Randrup and L.M. Satarov for stimulating discussions. 
This work was supported in part by EU-INTAS grant N\b{o} 94-3405. 
We are grateful to the Carlsberg Foundation and to the Rosenfeld Fund for financial support.

\end{document}